\documentclass[12pt]{iopart}

\usepackage[english]{babel} 
\expandafter\let\csname equation*\endcsname\relax
\expandafter\let\csname endequation*\endcsname\relax
\usepackage{amsmath,amsfonts,amssymb,amsthm,mathtools,physics}
\usepackage{hyperref}

\theoremstyle{definition}

\newcommand{\set}[1]{\left\{ {#1} \right\}}
\newcommand{\prt}[1]{\left( {#1} \right)}

\newcommand{\scal}[1]{\left< {#1} \right>}

\newcommand{\pd}[2]{\frac{\partial{#1}}{\partial{#2}}}

\newcommand{\setR}{{\mathbb R}}
\newcommand{\setC}{{\mathbb C}}

\newcommand{\lequi}{\ \ \Longleftrightarrow\ \ }

\newcommand{\D}{\mathcal{D}}

\renewcommand{\H}{\mathcal{H}}

\newcommand{\C}{\mathcal{C}}

\renewcommand{\bar}{\overline}		

\newcommand\bbone{{\mathbb{I}}}

\newcommand\fabeta{{\tilde{f}}}
\newcommand\gabeta{{\tilde{g}}}

\newcommand{\falpbeta}{{\check{f}}}

\newcommand{\bbbone}{{\text{\usefont{U}{dsss}{m}{n}\char49}}}

\begin{document}

\title{Quantum causality constraints on kappa-Minkowski space-time}

\author{Nicolas Franco$^{1,2}$ and Jean-Christophe Wallet$^3$}
\address{$^1$ Namur Institute for Complex Systems (naXys) and Department of Mathematics, University of Namur, Namur, Belgium}
\address{$^2$ Interuniversity Institute of Biostatistics and statistical Bioinformatics (I-BioStat) and Data Science Institute, University of Hasselt, Hasselt, Belgium}
\address{$^3$ IJCLab, Universit\'e Paris-Saclay, CNRS/IN2P3, 91405 Orsay, France}
\eads{\mailto{nicolas.franco@unamur.be}, \mailto{jean-christophe.wallet@universite-paris-saclay.fr}}

\vspace{10pt}

\begin{abstract}
The $\kappa$-Minkoswki space-time provides a quantum noncommutative-deformation of the usual  Minkowski space-time. However, a notion of causality is difficult to be defined in such a space with noncommutative time. In this paper, we define a notion of causality on a (1+1)-dimensional $\kappa$-Minkoswki space-time using the more general framework of Lorentzian noncommutative geometry. We show that this notion allows specific causal relations, but limited by a general constraint which is a quantum generalization of the traditional speed of light limit.
\end{abstract}

%
%
%
%
%

\section{Introduction}\label{section1}

It is needless to say that causality is an essential property of any physical theory or model. Some approaches to Quantum Gravity, under developments, are mainly based on causality, as building block. These approaches are causal-set \cite{causal-set} which mixes causality and discreteness, or the causal fermion systems \cite{causalfermion} which seem to reduce to the Standard Model and/or General Relativity in some suitably chosen limits. Both of these approaches belong to the commutative world and use basically a somehow standard description of causality.

However, it turns out that the notion of causality is not unique whenever a noncommutative geometry framework is assumed. For a review on different approaches to noncommutative (quantum) causality, see for instance \cite{besnard1,besnard2} and references therein. For a recent phenomenological review on possible tests of physics at the Planck scale including causality aspects, see \cite{review-quantgrav}.

Among these approaches, Lorentzian noncommutative geometry \cite{nicolas1} has been introduced as an attempt to accommodate Lorentzian signature in the by-now standard noncommutative geometry framework usually developed in Euclidean signature \cite{C94,MC08}. This natural evolution of noncommutative geometry is now arrived at a point where applications in fundamental physics are possible. One essential related feature is that Lorentzian noncommutative geometry can be linked to a well defined notion of causality \cite{nico1}, thus opening the possibility to equip a quantum (i.e.~noncommutative) space-time with causal structures. This type of noncommutative causality, which exactly reduces to the usual one at the commutative (low energy) limit, is connected to the Dirac operator acting as a metric. Recall that in the usual Euclidean set-up, the Dirac operator is the building block of the Connes spectral distance \cite{C94,MC08} extending the notion of geodesic distance in a noncommutative world. For explicit constructions on various noncommutative spaces, see \cite{spectrodist1,spectrodist2,spectrodist3}.

This notion of noncommutative causality has already been used to almost noncommutative manifolds \cite{nico2,nico22} for which the inherent causal structures give rise to a Zitterbewegung of a Dirac fermion \cite{nico-eck}. A somewhat similar analysis has been carried out for the Moyal plane equipped with a Minkowskian metric, sometimes called the "quantum Minkowski space-time" \cite{nicojc}, where it was shown that causal structures can actually exist in this quantum space, a point which has been subject to controversy in the area on (noncommutative) quantum field theories \cite{baletal,baletal2}, and by the way suggesting that, contrary to the common belief \cite{pheno1,pheno2}, causal structure need not breakdown at the Planck scale. In particular, it was shown that the causal structure occurring within a particular class of pure states is similar to the causal structure on the usual Minkowski space, while the notion of locality is lost.\\

The purpose of this paper is to present a first exploration of the above quantum causality on $\kappa$-Minkowski space-time \cite{majid-ruegg,LUKIERSKI1991331}, which is one of the most studied noncommutative spaces and an interesting candidate in the developpement of some Quantum Gravity. $\kappa$-Minkowski space-time can be defined as the dual of a subalgebra of the $\kappa$-Poincar\'e algebra involving the so-called deformed translation, and may be viewed as the universal enveloping algebra of the Lie algebra of coordinates $[x_0,x_i]=\frac{i}{\kappa}x_i$, $[x_i,x_j]=0$, under a deformation parameter $\kappa>0$. Studying different notions of quantum causality which may be relevant at the Planck scale is far from being of a purely academic interest. For instance, it is worth pointing out that current experiments on in-vacuo dispersion relations of very highly energetic cosmological photons might be definitely unable to detect signals stemming from some $\kappa$-deformation if causality at the Planck scale is of the type proposed in \cite{mercati} whereas observation might eventually remain possible if the usual "classical" causality at the Planck scale holds true.

 In this paper, we show that a valid causal structure can be defined on an interesting set of pure states on $\kappa$-Minkowski space-time, here limited to the (1+1)-dimensional case. While it is difficult to have an explicit characterisation of the entire causal structure, we are able to extract some specific causality constraints. In particular, we show that there exists some specific sufficient conditions of allowed causal relations corresponding to a phase-momentum transport between two quantum states. Then we highlight an interesting necessary constraint, hence a clear limit of causal evolution between quantum states, which can be interpreted as a quantification of the classical speed of light limit.\\

The paper is organized as follows:
\begin{itemize}
    \item The first part (Section \ref{section2}) describes the necessary basic elements to introduce the $\kappa$-Minkowski space-time and the construction of a Lorentzian spectral triple on it: in Section \ref{section21}, we present the general definition of a Lorentzian spectral triple; in Section  \ref{section22} the algebra of the spectral triple and the related Hilbert space; and in Section \ref{section23} the associated Dirac operator.
 \item The second part (Section \ref{section3}) presents a first exploration of the notion of causality on $\kappa$-Minkowski space-time with particular results: in Section \ref{section31}, we present and apply the notion of causality to $\kappa$-Minkowski yielding to a general unsolved condition; in Section \ref{section32} a particular solution of the general condition is presented as a sufficient causality condition corresponding to a phase-momentum transport; and in Section \ref{section33} we present another specific solution giving rise to a necessary quantum causality constraint.
\end{itemize}

\section{Quantum space-time and Lorentzian spectral triple}\label{section2}
\subsection{Basic features of Lorentzian spectral triple}\label{section21}

The main tool used within this paper is the so called Lorentzian spectral triple \cite{nicolas1}. Such a spectral triple with Lorentzian signature is not very different from its Riemannian counterpart \cite{C94,MC08}, except that the Dirac operator is naturally self-adjoint in a Krein space \cite{stroh}. Recall that a Krein space is a space with indefinite inner product, supplemented by additional conditions \cite{bog}. This space replaces the Hilbert space which is one of the main data defining a standard spectral triple. However, a Hilbert space can still be used in the definition of a Lorentzian spectral triple thanks to the introduction of a specific operator $\mathcal{J}$, called the fundamental symmetry{\footnote{In the case of "Minkowskian Moyal plane" mentioned above, $\mathcal{J}=i\gamma^0$, the time-like Dirac gamma matrice.}} which permits one to trade the Krein space for a Hilbert space (and vice versa). Different definitions of Lorentzian spectral triples appeared in the literature, e.g. \cite{lrtz-def,lrtz-def2} which however are compatible. In the sequel, we will use an adapted case of these latter \cite{nico-tempo} whose advantage is that no signature other than the Lorentzian one is allowed so a notion of causality is always well defined.\\

We will use as general definition for a Lorentzian spectral triple the set of data:
\begin{equation}
\big\{\mathbb{A},\ \widetilde{\mathbb{A}},\ \pi,\ \mathcal{H},\ D,\ \mathcal{J}  \big\}\label{data1},
\end{equation}
where $\mathbb{A}$ is a non-unital pre-C$^*$ algebra with a preferred unitalization (still a pre-C$^*$ algebra) $\widetilde{\mathbb{A}}$ with $\mathbb{A}$ as an ideal, while $\pi$ is a faithful $*$-representation of $\mathbb{A}$ on $\mathcal{B}(\mathcal{H})$, the algebra of bounded operators on a Hilbert space $\mathcal{H}$, $\pi:\mathbb{A}\to\mathcal{B}(\mathcal{H})$ which also acts as a faithful $*$-representation of $\widetilde{\mathbb{A}}$ on $\mathcal{B}(\mathcal{H})$. In \eqref{data1}, $D$ is the unbounded Dirac operator with dense domain $\text{Dom}(D)$ in $\mathcal{H}$ satisfying
\begin{equation}
\forall a\in\widetilde{\mathbb{A}},\ [D,\pi(a)]\in\mathcal{B}(\mathcal{H}),\label{cond1}
\end{equation}
\begin{equation}
\forall a\in{\mathbb{A}},\ \pi(a)(1+\langle D\rangle^2)^{-\frac{1}{2}}\in\mathcal{K}(\mathcal{H}),\ \text{with}\ \langle D\rangle^2:=\frac{1}{2}(D^\dag D+DD^\dag),\label{cond1bis}
\end{equation}
where $\mathcal{K}(\mathcal{H})$ denotes the algebra of compact operators on $\mathcal{H}$ and $\dag$ is the involution of $\mathbb{A}$. \\
The operator $\mathcal{J}\in\mathcal{B}(\mathcal{H})$ related to the fundamental symmetry must satisfy
\begin{equation}
\mathcal{J}^2=1,\ \ \mathcal{J}^\dag=\mathcal{J},\ \ [\mathcal{J},\pi(a)]=0,\ \forall a\in\widetilde{\mathbb{A}}\label{cond2},
\end{equation}
together with
\begin{equation}
D^\dag\mathcal{J}=-\mathcal{J}D\label{cond3}
\end{equation}
which must hold true on $\text{Dom}(D)=\text{Dom}(D^\dag)$.\\
Finally, the conditions \eqref{cond1}-\eqref{cond3} are supplemented by the additional condition that there exists a densely-defined self-adjoint operator $\mathcal{T}$ and a positive element $N\in\widetilde{\mathbb{A}}$ such that $\text{Dom}(D)\cap\text{Dom}(\mathcal{T})$ is dense in $\mathcal{H}$ and 
\begin{equation}
(1+\mathcal{T})^{-\frac{1}{2}}\in\widetilde{\mathbb{A}}\label{cond5},
\end{equation}
\begin{equation}
\mathcal{J}=-N[D,\mathcal{T}]\label{cond6}.
\end{equation}
At this point, some comments are in order.
\begin{itemize}
\item[i)] As already mentioned above, $\mathcal{J}$ corresponds to the fundamental symmetry connecting Hilbert space and Krein space. If $\langle.\ ,\ .\rangle$ denotes the positive definite product equipping a Hilbert space $\mathcal{H}$, then the corresponding indefinite inner product on the corresponding Krein space is given by $(.\ ,\ .)_\mathcal{J}=\langle\ .\ , \mathcal{J}\ .\rangle$.
\item[ii)] According to the above comment, \eqref{cond3} expresses the fact that the operator $iD$ is self-adjoint w.r.t the indefinite inner product $(.\ ,\ .)_\mathcal{J}$.
\item[iii)] We note that the resolvent condition of the standard (Riemannian) spectral triple{\footnote{ Recall that this latter states that for any $a\in\mathbb{A}$ and any $\lambda$ {\it{not}} in the spectrum of $D$, one has $\pi(a)( D-\lambda)^{-1}\in\mathcal{K}( \mathcal{H})$.}} does not appear in the definition of a Lorentzian spectral triple, stemming from the fact that $D$ is no longer an elliptic operator in this non-Riemannian context. This explains the appearance of $\langle D\rangle^2:=\frac{1}{2}(D^\dag D+DD^\dag)$ in \eqref{cond1bis} leading to an elliptic operator $(1+\langle D\rangle^2)^{\frac{1}{2}}$, self-adjoint for the Krein product $(.\ ,\ .)_\mathcal{J}$, whose inverse \eqref{cond1bis} must be compact \cite{stroh}.
\item $\mathcal{T}$ in \eqref{cond5}, \eqref{cond6} can be viewed as a global time function for the Lorentzian spectral triple \cite{nico-tempo}.
\item The unital algebra $\widetilde{\mathbb{A}}$ serves for technical purpose only. It could not support a notion of causality since a unital algebra corresponds to a compact space on which no causality can be defined. Notice that the Riemannian spectral triple for the Moyal plane \cite{ioch} also uses a preferred unitalization of the non unital algebra modeling the Moyal plane. 
\end{itemize}
\subsection{Lorentzian spectral triple for $\kappa$-Minkowski space-time: the algebra}\label{section22}

In this paper, we will consider the $(1+1)$-dimensional version of the $\kappa$-Minkowski space-time. It can be defined \cite{majid-ruegg} as the dual of a subalgebra of the $\kappa$-Poincar\'e algebra involving the so-called deformed translations. See \ref{apendixA} for useful formulas. Its simplest presentation is provided by the universal enveloping algebra of the following solvable Lie algebra of coordinates
\begin{equation}
[x_0,x]=\frac{i}{\kappa}x\label{kappaminko},
\end{equation}
where $\kappa>0$ is the deformation parameter. The corresponding, hence solvable, Lie group denoted by $\mathcal{G}$ is known to be the orientation-preserving affine group of the real line, see e.g. \cite{groupaff2}, \cite{groupaff1}, namely
\begin{equation}
\mathcal{G}=\mathbb{R}\ltimes_\phi\mathbb{R}\label{affinegroup}.
\end{equation}
It is a simply connected Lie group which is non unimodular, with group product, inverse and unit respectively defined by the following map $W:\mathbb{R}\times\mathbb{R}\to\mathcal{G}$
\begin{eqnarray}
W(p_0,p_1)W(q_0,q_1)&=&W(p_0+q_0,p_1+e^{-p_0/\kappa}q_1),\label{grouplaw1}\\
W^{-1}(p_0,p_1)&=&W(-p_0,-e^{p_0/\kappa}p_1),\label{grouplaw2}\\
\bbone_{\mathcal{G}}&=&W(0,0)\label{grouplaw3},
\end{eqnarray}
while the automorphism $\phi$ in \eqref{affinegroup} is defined for any $q\in\mathbb{R}$ by
\begin{equation}
\phi:\mathbb{R}\to\text{Aut}(\mathbb{R}),\ \ \phi(p_0)q=e^{-p_0/\kappa}q\label{automorph}.
\end{equation}
\\
A suitable characterization of the associative algebra modeling the quantum (i.e.~noncommutative) $\kappa$-Minkowski space-time can be obtained \cite{DubSit}, \cite{PW2018}, \cite{matassa} by starting from the convolution algebra of $\mathcal{G}$, denoted by $(L^1(\mathcal{G}),d\rho)$, where $d\rho$ is some Haar measure on $\mathcal{G}$ to be fixed in a while, which is (isomorphic to) the completion 
w.r.t. the $L^1$-norm $||F||_1=\int_{\mathcal{G}}F(u)d\rho(u)$ of the algebra involving the $\mathbb{C}$-valued continuous functions on $\mathcal{G}$ with compact support $\mathcal{C}_c(\mathcal{G})$. The resulting algebra $(L^1(\mathcal{G}),d\rho)$ is a Banach $*$-algebra with has a natural involution defined by 
\begin{equation}
F^\ddag(u)={\bar{F}}(u^{-1})\Delta(u^{-1})\label{invol-banach},
\end{equation}
for any $F\in (L^1(\mathcal{G}),d\rho)$, where $\bar{F}$ denotes the complex conjugate of $F$ and the group homomorphism $\Delta:\mathcal{G}\to\mathbb{R}^+$ is the modular function linking together the left-invariant and right-invariant Haar measures of $\mathcal{G}$, respectively denoted by $d\mu$ and $d\nu$, through the well known relation
\begin{equation}
d\nu(u)=\Delta(u^{-1})d\mu(u)\label{modular-relation},
\end{equation}
leading to the useful formulas
\begin{equation}
d\nu(u^{-1})=\Delta(u)d\nu(u),\ \ d\mu(u^{-1})=\Delta(u^{-1})d\mu(u)\label{relations-bis}.
\end{equation}
In the present situation, the modular function is given by
\begin{equation}
\Delta(u)=\Delta(p_0,p_1)=e^{\frac{p_0}{\kappa}}\label{deltaegal}.
\end{equation}
The convolution product equipping the convolution algebra can be expressed in terms of either the left -invariant Haar measure $d\mu$ or alternatively its right-invariant counterpart $d\nu$ as given below
\begin{equation}
(F\circ G)(s)=\int_{\mathcal{G}}F(us)G(s^{-1})d\mu(s)=\int_{\mathcal{G}}F(us^{-1})G(s)d\nu(s)\label{convolution}.
\end{equation}
Note that $(L^1(\mathcal{G}),d\mu)$ and $(L^1(\mathcal{G}),d\nu)$ are isometrically isomorphic through the following map $\omega:C_c(\mathcal{G})\to C_c(\mathcal{G})$, $(\omega(F))(u)=F(u^{-1})$. The above properties are standard results form harmonic analysis of locally compact groups.\\

In what follows, we will use the formulation based on the left-invariant Haar measure. This later is given by
\begin{equation}
d\mu(u)=d\mu(p_0,p_1)=e^{\frac{p_0}{\kappa}}dp_0dp_1\label{left-haar}.
\end{equation}
By combining \eqref{left-haar} with \eqref{grouplaw1} and \eqref{grouplaw2}, the convolution product and the involution \eqref{invol-banach} can then be cast into the form
\begin{equation}
(F\circ G)(p_0,p_1)=\int\ dq_0dq_1e^{\frac{q_0}{\kappa}}\ F(q_0,q_1)G(p_0-q_0,e^{\frac{q_0}{\kappa}}(p_1-q_1))\label{convol-fin},
\end{equation}
\begin{equation}
F^\ddag(p_0,p_1)={\bar{F}}(-p_0,-p_1e^{\frac{p_0}{\kappa}})e^{-\frac{p_0}{\kappa}}\label{invol-fin}.
\end{equation}
A suitable associative algebra modeling the $\kappa$-deformation which can be used to define the Lorentzian spectral triple for $\kappa$-Minkowski space-time can be determined from the characterization of the group C*-algebra for $\mathcal{G}$, denoted by $C^*(\mathcal{G})$. One first observes that since $\mathcal{G}$ is a solvable group, it is amenable, thus one has $C^*(\mathcal{G})=C_{red}^*(\mathcal{G})$ where $C_{red}^*(\mathcal{G})$ is the reduced algebra obtained by completing $ (L^1(\mathcal{G}),d\mu)$ with the left regular representation on $(L^2(\mathcal{G}),d\mu)$.
For our purpose, il will be sufficient to restrict ourselves to its generating dense subalgebra of smooth functions on $\mathcal{G}$ with compact support, denoted by $\mathcal{C}^\infty_c(\mathcal{G})$. \\
Now from standard results on C*-algebras of crossed products of groups \cite{groupaff1}, one infers that $C^*(\mathcal{G})=C^*(\mathbb{R}\ltimes\mathbb{R})\simeq\mathbb{R}\ltimes C^*(\mathbb{R})$ generated by the dense subalgebra of smooth functions of variables $p_0$ and $p_1$, having compact support for $p_0$ and taking their values in the set of smooth functions of $p_1$ with compact support. We denote this algebra by
\begin{equation}
\mathcal{C}^\infty_c(\mathcal{G}):=\mathcal{C}^\infty_c(\mathbb{R}, \mathcal{C}^\infty_c(\mathbb{R})).
\end{equation}
\\
From now on, we choose this algebra, denoted by $\mathbb{A}_p$ as the starting algebra for the spectral triple, whenever the functions depend on the variables $p_0,p_1$, interpreted from now on as momenta. Namely, one has
\begin{equation}
\mathbb{A}_p:=\mathcal{C}^\infty_c(\mathcal{G})\label{algebrenp}.
\end{equation}
In order to make contact with the physics literature, we further interpret any of the above functions $F(p_0,p_1)$ as the Fourier transform of some function $f(x_0,x_1)$ where $x_0,x_1$ are space-time variables.
Namely, $F(p_0,p_1)=\mathcal{F}f(p_0,p_1)=\int d^2x\ e^{-i(p_0x_0+p_1x_1)}f(x_0,x_1)$. As a consequence of the Paley-Wiener theorem, the algebra $\mathbb{A}_p$ is isomorphically mapped by inverse Fourier transform into the algebra $\mathbb{A}_x$ of functions which are analytic and of exponential type in the variable $x_0$ with values in the space of analytic functions of exponential type in the variable $x_1$. Namely
\begin{equation}
  \mathbb{A}_x:=\mathcal{F}^{-1}A_p=\mathfrak{E}(\mathbb{R},\mathfrak{E}(\mathbb{R}))\label{algebrenx},
\end{equation}
where $\mathfrak{E}(\mathbb{R})$ is the set of functions whose analytic continuation  $f:z\in\mathbb{C}\to f(z)$ is an entire function  on $\mathbb{C}$ verifying the "exponential" bound $|f(z)|\le K_1e^{K_2|\Im{z}|}$, with $K_1>0,\ K_2>0$.
The product equipping $\mathbb{A}_x$, of course related to the product \eqref{convol-fin}, is defined by
\begin{equation}
\mathcal{F}(f\star g)=\mathcal{F}f\circ\mathcal{F}g\label{star-pro}.
\end{equation}
A simple calculation yields
\begin{equation}
(f\star g)(x_0,x_1)=\int \frac{dp_0}{2\pi}dy_0\ e^{-iy_0p_0}f(x_0+y_0,{x}_1)g(x_0,e^{-p_0/\kappa}{x}_1)  \label{starpro-2d}.
\end{equation}
In the same way, the involution \eqref{invol-fin} translates into
\begin{equation}
    f^\dag(x_0,x_1)= \int \frac{dp_0}{2\pi}dy_0\ e^{-iy_0p_0}{\bar{f}}(x_0+y_0,e^{-p_0/\kappa}{x}_1)\label{invol-2d}.
\end{equation}

Three remarks are in order here:
\begin{itemize}
\item Notice that \eqref{starpro-2d} and \eqref{invol-2d} are nothing but the star-product and involution used in \cite{DubSit}, \cite{PW2018}, \cite{matassa} to describe the $\kappa$-deformation of the Minkowski space-time. This latter was obtained from a mere Weyl quantization. Recall that the corresponding quantization map is a morphism of $*$-algebra defined by $Q(f)=\pi(\mathcal{F}f)$ where the non-degenerate unitary $*$-representation $\pi:L^1(\mathcal{G},d\mu)\to {\mathcal{B}}(\mathcal{H})$, with $\mathcal{H}$ some Hilbert space to be defined in a while, is given by $\pi(f)=\int_{\mathcal{G}}d\mu(s)f(s)\pi_U(s)$, where $\pi_U:\mathcal{G}\to{\mathcal{B}}(\mathcal{H})$ denotes some irreducible unitary representation of $\mathcal{G}$. Simply writing $Q(f\star g)=Q(f)Q(g)=\pi(\mathcal{F}(f))\pi(\mathcal{F}(g))=\pi(\mathcal{F}(f)\circ\mathcal{F}(g))$ yields \eqref{star-pro} leading to \eqref{starpro-2d}. A similar argument yields \eqref{invol-2d}.
\item The unitary irreducible representations of $\mathcal{G}=\mathbb{R}\ltimes\mathbb{R}$, $\pi_U:\mathcal{G}\to\mathcal{H}=L^2(\mathbb{R},ds)$, have been classified in \cite{gelfand1} (see also \cite{groupaff2}). The only non-trivial one are defined by
\begin{equation}
    (\pi_{U\pm}(p_0,p_1)\phi)(s)=e^{\pm ip_1e^{-s}}\phi(s+p_0),\label{gelf1}
\end{equation}
for any $\phi\in\mathcal{H}$ while all the other unitary irreducible one are 1-dimensional. It follows that the non-trivial unitary representations of the corresponding group algebra of $\mathcal{G}$ are given by
\begin{equation}
(\pi_\pm(f)\phi)(s)=\int\ dp_0dp_1e^{p_0-s}\mathcal{F}f(p_0-s,p_1)e^{\pm ip_1e^{-s}}\phi(u),\label{decadix}
\end{equation}
while the trivial 1-dimensional one is
\begin{equation}
  (\pi_0(f)\phi)(s)=\int\ dp_0dp_1e^{p_0-s}\mathcal{F}f(p_0-s,p_1)\phi(u).\label{decadix2}
\end{equation}
Expressed in the spatial variables, \eqref{decadix} takes the form
\begin{equation}\label{reprerpnu}(\pi_{\pm}(f) \,\phi)(s)  = \tfrac{1}{2\pi} \int \dd u \dd v\; f(v, \pm e^{-s}) \,e^{-i v (u-s)}\, \phi(u),
\end{equation}
while \eqref{decadix2} takes the compatible degenerate form
\begin{equation}\label{reprerpnuzero}(\pi_{0}(f) \,\phi)(s)  = \tfrac{1}{2\pi} \int \dd u \dd v\; f(v, 0) \,e^{-i v (u-s)}\, \phi(u).
\end{equation}
Recall also that 
\begin{equation}
\hat{\pi}=\pi_+\oplus\pi_-\label{faithful}
\end{equation}
is a faithful representation.
\item In the course of the analysis, we will need to work with functions depending on mixed variables $(p_0, x_1)$ obtained by partial Fourier transform. The relevant algebra becomes 
\begin{equation}
\mathbb{A}_{px}=\mathcal{C}_c^\infty(\mathbb{R}, \mathfrak{E}(\mathbb{R})),\label{algebrenpx}
\end{equation}
stemming from the Paley-Wiener theorem. It is a straightforward calculation to obtain the product between functions in the above mixed variables, denoted by $\tilde{f}(p_0,x_1)$. It is given by
\begin{equation}
(\fabeta \star \gabeta)(p_0,x_1) = \int \dd q_0\; \fabeta(q_0,x_1)\, \gabeta(p_0-q_0, e^{-q_0/\kappa} x_1).
\end{equation}
Under this mixed formalism, the representations \eqref{reprerpnu}-\eqref{reprerpnuzero} take the form
\begin{equation}(\pi_{\nu}(\fabeta) \,\phi)(s) = \int \dd u\; \fabeta(u-s, \nu e^{-s})\, \phi(u),\label{reprerpmixed} \end{equation}
where $\nu$ is used to represent elements in $\set{+,0,-}$.
\end{itemize}

At this stage, we have characterized above a suitable algebra for the Lorentzian spectral triple, expressed by \eqref{algebrenp}, \eqref{algebrenx}, \eqref{algebrenpx} according to the choice of variables for the functions. Set now
\begin{equation}
\mathcal{H}_{+,0,-}:=(L^2(\mathbb{R}),ds)\otimes\mathbb{C}^2.\label{blockhilbert}
\end{equation}
Since we will deal with a 2-dimensional Dirac operator, a suitable choice for the Hilbert space involved in the triple is
\begin{equation}
\mathbb{H}=\mathcal{H}_{+}\oplus\mathcal{H}_{0}\oplus\mathcal{H}_{-}\label{hilbert-final}.
\end{equation}
Besides, keeping in mind \eqref{faithful}, a suitable choice for the faithful representation entering the spectral triple is
\begin{equation}
\Pi=(\pi_+\oplus\pi_0\oplus\pi_-)\otimes\bbone_2.\label{cestpi}
\end{equation}

As recalled in the subsection \ref{section21}, the algebra involved in the spectral triple must be supplemented by an unitized counterpart. We will denote is by $\tilde{\mathbb{A}}$ with the same subscripts than those used for $\mathbb{A}$ according to the used set of variables. We will choose the multiplier algebra of $\mathbb{A}$, $M(\mathbb{A})$ characterized in \cite{DubSit}. Recall that $M(\mathbb{A})_x$ is the algebra of smooth functions with polynomial bounds whose derivatives are also with polynomial bounds. It involves in particular the coordinate functions and the constants. Expressed in the different set of variables, the unit function is expressed as
\begin{equation}
\bbbone(p_0,p_1)=(\delta(p_0),\delta(p_1)),\ \  \bbbone(p_0,x_1)=\delta(p_0),\ \ \bbbone(x_0,x_1)=1\label{unities},
\end{equation}
respectively corresponding to the unit of $\tilde{\mathbb{A}}_p$, $\tilde{\mathbb{A}}_{px}$ and $\tilde{\mathbb{A}}_x$. The mixed variable unitalization $\tilde{\mathbb{A}}_{px}$, which we will use later on, contains functions with compact support on the first variable $p_0$. Note that $\Pi$ \eqref{cestpi} extends to a faithful representation of $\tilde{\mathbb{A}}$ on $\mathcal{B}(\mathbb{H})$.

\subsection{Lorentzian spectral triple for $\kappa$-Minkowski space-time: the Dirac operator}\label{section23}
In the subsequent analysis, we will consider the Dirac operator obtained from natural one-parameter groups of automorphisms of ${C}^*(\mathcal{G})$, which has already been considered in \cite{IMS}. It is known \cite{groupaff1} that ${C}^*(\mathcal{G})$ has a natural group of automorphisms defined by the modular group of $*$-automorphisms $t\in\mathbb{R}\mapsto\sigma_t\in\text{Aut}({C}^*(\mathcal{G}))$,
\begin{equation}
\sigma_t(f)=e^{iP_0/\kappa}\triangleright f=e^{t\partial_0/\kappa}\triangleright f\label{sigmat}
\end{equation}
for any $f\in\mathbb{A}_x$, in the notations of \cite{PW2018}. It follows that $\kappa\frac{d\sigma_t}{dt}|_{t=0}\triangleright f$ for any $f\in\mathbb{A}_x$ defines (the action of) a derivation which is simply $D_0=\partial_0$.\\
Another group of automorphisms can be obtained by making use of Fourier transform in
$C^*(\mathcal{G})=\mathbb{R}\ltimes C^*(\mathbb{R})$. Indeed, by Fourier transform, one has $C^*(\mathbb{R})\simeq C_0(\mathbb{R})$, the $C^*$-algebra of continuous functions vanishing at infinity. Hence $C^*(\mathcal{G})=\mathbb{R}\ltimes C_0(\mathbb{R}) $, generated by $\mathbb{A}_{px}$, with action of $\mathbb{R}$ on $C_0(\mathbb{R})$ given by $\omega_{p_0}(f)(x_1)=f(e^{-p_0/\kappa}x_1)$ for any $f\in C_0(\mathbb{R})$. This therefore corresponds to the group of automorphisms $t\in\mathbb{R}\mapsto\omega_{-t}\in\text{Aut}(C_0(\mathbb{R}))$, extending to $t\in\mathbb{R}\mapsto\omega_t\in\text{Aut}({C}^*(\mathcal{G}))$ with
\begin{equation}
    \omega_t(\tilde{f})(p_0,x_1)=\tilde{f}(p_0,e^tx_1)
\end{equation}
for any $f\in\mathbb{A}_{px}$ which gives also $\omega_t({f})(x_0,x_1)={f}(x_0,e^tx_1)$ for any $f\in\mathbb{A}_x$. Thus, the second derivative is defined by $\frac{d\omega_t(f)}{dt}|_{t=0}$ which corresponds to the derivation $D_1=x_1\partial_1$.\\
At this point, two comments are in order:
\begin{itemize}
\item It turns out that the derivation $D_1$ is inner. Indeed, a simple computation yields $(x_0\star f)(x_0,x_1)=x_0f(x_0,x_1)+\frac{i}{\kappa}x_1\partial_1 f(x_0,x_1)$ and $(f\star x_0)(x_0,x_1)=x_0f(x_0,x_1)$ where we used \eqref{starpro-2d}. Therefore, one obtains
\begin{equation}
ix_1\partial_1f=\kappa[x_0,f]
\end{equation}
for any $f\in\mathbb{A}_x$. Hence the result. The derivations $D_0,D_1$ are obviously real derivations and obeying an untwisted Leibnitz rule. Note however that $D_1$ vanishes at the commutative limit $\kappa\to\infty$. These derivations have been used to build the Dirac operator underlying \cite{IMS} as an attempt to obtain a regular spectral triple for a two-dimensional $\kappa$-deformation of (Euclidean version of) the Minkowski space, thus working in the Euclidean set-up of the standard spectral triples. \\
We will consider such a Dirac operator in the following and study causal structures arising in this $\kappa$-deformation, thus initiating a first exploration of the causality in $\kappa$-deformed Minkowski space.
\item Notice that other choices for the derivations are possible. In particular, one could start from the twisted derivations used in \cite{kappagauge1}, \cite{kappagauge2} (see also \cite{matassa}) to construct a $\kappa$-Poincar\'e invariant gauge theory on $\kappa$-Minkowski space-time which can be viewed as a noncommutative analog of a Yang-Mills theory. These are given by 
\begin{equation}
 X_0=\kappa\mathcal{E}^\gamma(1-\mathcal{E}),\ \ X=\mathcal{E}^\gamma P\label{tausig-famil}
\end{equation}
with $[X_0,X]=0$, where $\gamma\in\mathbb{R}$. The derivations \eqref{tausig-famil} obviously reduce to the usual derivations at the commutative limit $\kappa\to\infty$. However, they obeys a twisted Leibnitz rule given by
\begin{equation}
X_\mu(f\star g)=X_\mu(f)\star (\mathcal{E}^\gamma\triangleright g)+ (\mathcal{E}^{1+\gamma}\triangleright f)\star X_\mu(g),\label{tausigleibniz}
\end{equation}
for any $f,g\in\mathbb{A}_x$ where the action of $\mathcal{E}$ and $P$ on the algebra is given by \eqref{left-module1}. Observe that the derivations \eqref{tausig-famil} are {\it{not}} real derivations since the following identity holds true
\begin{equation}
(X_\mu(f))^\dag=-\mathcal{E}^{-2\gamma-1}\triangleright (X_\mu(f^\dag))\ne X_\mu(f^\dag), \label{laformule}
\end{equation}
contrary to the above derivations $D_0,D_1$. The causal structures stemming from a Dirac operator built from the derivations \eqref{tausig-famil} will be examined in a future publication. \\
\end{itemize}

As it is now clear that the derivation corresponding to the second spatial variable $D_1=x_1\partial_1$ vanishes as $\kappa\to\infty$ and that the non-trivial unitary representations $\pi_\pm$ \eqref{reprerpnu} can only be distinguished from the trivial one $\pi_0$ \eqref{reprerpnuzero} by this second variable, we have that our spectral triple is inconsistent and collapses as $\kappa\to\infty$. For the simplicity of future expressions, we will work from now with a fixed finite $\kappa=1$ as others finite values are pure rescaling.\\

We set now
\begin{equation}
    \partial_\pm=\partial_0\pm x_1\partial_1.\label{deltaplusmoins}
\end{equation}
The Dirac operator entering the Lorentzian spectral triple is defined as
\begin{equation}
  \D =-\gamma^k \partial_k \otimes \bbbone_3 =-i \left(\begin{matrix}0 &\partial_-\\\partial_+ &0\end{matrix}\right) \otimes \bbbone_3:=D\otimes\bbbone_3\label{diracoper},
\end{equation}
where the Dirac gamma matrices are
\begin{equation}
 \gamma^0 = \left(\begin{matrix}0 & i\\i&0\end{matrix}\right),\ \  \gamma^1 = \left(\begin{matrix}0 & -i\\i&0\end{matrix}\right). 
\end{equation}

The domain of $D$ is $\text{Dom}(D)=\mathcal{S}(\mathbb{R})\otimes\mathbb{C}^2$ which is obviously dense in each of the Hilbert spaces defined in \eqref{blockhilbert}. Hence $\text{Dom}(\mathcal{D})=\mathcal{S}(\mathbb{R})\otimes\mathbb{C}^6$.\\
The derivations $\partial_\pm$ \eqref{deltaplusmoins} act on the algebra. This can be translated as an action of the derivations on the Hilbert space \eqref{hilbert-final} which can be determined from the following action on $L^2(\mathbb{R},ds)$
\begin{equation}
(\partial_0 \phi)(s) =  s\, \phi(s),\quad
(\partial_1 \phi)(s) =  -i (\partial_s \phi)(s),\label{actionderivation}
\end{equation}
for any $\phi\in L^2(\mathbb{R},ds)$. \\

The fundamental symmetry $\mathcal{J}\in\mathcal{B}(\mathbb{H})$ is given by
\begin{equation}
    \mathcal{J}=i\gamma^0 \otimes \bbbone_3 \label{symfundam}.
\end{equation}
It is a simple matter of algebra to verify that $ \mathcal{J}^2=\bbone$, $ \mathcal{J}^\dag= \mathcal{J}$ and $[\mathcal{J},\Pi(a)]=0$ for any $a$ in the algebra, thus consistent with \eqref{cond3}.\\

We set
\begin{equation}
    \langle \Phi,\Psi \rangle=\sum_{\nu=+,-,0}\langle \varphi^{(\nu)},\psi^{(\nu)} \rangle_{\mathcal{H}_\nu}\label{hilbertproduct},
\end{equation}
 for any $\Phi,\Psi\in\mathbb{H}$ with $\Phi=\oplus_\nu\varphi^{(\nu)}$ according to the decomposition of $\mathbb{H}$ \eqref{hilbert-final} (and a similar decomposition for $\Psi$), where $\langle .,.\rangle_{\mathcal{H}_\nu}$, $\nu=+,-,0$ is the Hilbert product on $L^2(\mathbb{R})\otimes\mathbb{C}^2$ given by 
\begin{equation}
\langle \varphi^{(\nu)},\psi^{(\nu)} \rangle_{\mathcal{H}_\nu}=\int ds\ {\varphi^{(\nu)}}^\dag(s)\psi^{(\nu)}(s)\label{prodhilbertutile}.
\end{equation}
The Krein product related to $\mathcal{J}$ is defined by
\begin{equation}
    (\Phi,\Psi)_{\mathcal{J}}:=\langle \Phi,\mathcal{J}\Psi \rangle=\sum_{\nu=+,-,0}\langle \varphi^{(\nu)},i\gamma^0\psi^{(\nu)} \rangle_{\mathcal{H}_\nu}\label{kreinproduit}.
\end{equation}
It can be easily verified that one has
\begin{equation}
   (\Phi,\mathcal{D}\Psi)_{\mathcal{J}}= (\mathcal{D}\Phi,\Psi)_{\mathcal{J}}
\end{equation}\label{self-adj1}
and thus
\begin{equation}
   \mathcal{D}^\dag\mathcal{J}=-\mathcal{J}\mathcal{D} \label{self-adj2},
\end{equation}
which agrees with \eqref{cond2}.\\

Set now
\begin{equation}
\mathcal{T}=-i\ \oplus_\nu(\pi_\nu(x_0)\otimes\bbone_2)\label{timefunction}
\end{equation}
with domain $\text{Dom}(\mathcal{T})=\text{Dom}(\mathcal{D})$. The operator $\mathcal{T}$ is self-adjoint which simply  stems from 
\begin{equation}
(\pi_\nu(x_0)\varphi)(s)=-i\partial_s\varphi(s),
\end{equation}
for any $\varphi\in L^2(\mathbb{R})$, obtained from \eqref{reprerpnu}. From \eqref{timefunction}, one obtains the following blockwise decomposition
\begin{equation}
[\mathcal{D},\mathcal{T}]=-i\oplus_\nu[D,\pi_\nu(x_0)\otimes\bbone_2],
\end{equation}
so that one has to compute $[D,\pi_\nu(x_0)\otimes\bbone_2]$. Indeed, by using \eqref{reprerpnu} and \eqref{actionderivation}, one infers
\begin{eqnarray}
(\partial_0\pi_\nu(a)\varphi)(s)&=&\int dudv\ sf(v,e^{-s}\nu)e^{-iv(u-s)}\varphi(u)\label{formdezero},\\
(\pi_\nu(a)\partial_0\varphi)(s)&=&\int dudv\ f(v,e^{-s}\nu)e^{-iv(u-s)}u\varphi(u)\label{formdun},
\end{eqnarray}
for any $a\in\mathbb{A}_x$, $\varphi\in L^2(\mathbb{R})$ and $\nu=+,-,0$. From \eqref{formdezero} and \eqref{formdun}, one easily obtains
\begin{equation}
[\partial_0,\pi_\nu(a)]=i\pi_\nu(\partial_0 a)\label{commutzero}
\end{equation}
for any $a\in\mathbb{A}_x$ and $\nu=+,-,0$. A similar computation yields
\begin{equation}
  [\partial_1,\pi_\nu(a)]=i\pi_\nu(\partial_1 a)\label{commutun}, 
\end{equation}
for any $a\in\mathbb{A}_x$ and $\nu=+,-,0$. \eqref{commutzero} and \eqref{commutun} imply
\begin{equation}
[\partial_0,\pi_\nu(x_0)]=i,\ \ [\partial_1,\pi_\nu(x_0)]=0\label{commutxzero}
\end{equation}
which gives rise to
\begin{equation}
[\mathcal{D},\mathcal{T}]=-\mathcal{J},
\end{equation}
which fulfills \eqref{cond6} provided $N$ is the identity operator while the self-adjoint operator $\mathcal{T}$ \eqref{timefunction} verifies the condition \eqref{cond5}.\\

Furthermore, since $\pi_\nu(a)$ is bounded for any $a\in\mathbb{A}_x$, it follows from \eqref{commutzero} and \eqref{commutun} that the operators $[\partial_k,\pi_\nu(a)]$, $k=0,1$, are bounded operators on $L^2(\mathbb{R})$ for any $a\in\mathbb{A}_x$. Hence
\begin{equation}
\forall a\in{\mathbb{A}_x},\ [\mathcal{D},\Pi(a)]\in\mathcal{B}(\mathbb{H}),\label{commutbounded}
\end{equation}
which extend to any $a\in\tilde{\mathbb{A}}_x$, thus verifying the condition \eqref{cond1}.\\

Let us consider the compactness condition \eqref{cond1bis}. First, one easily realises that $\Pi(a)(1+\mathcal{D}^2)^{-1/2}$ for any $a\in\mathbb{A}$, with $\Pi$ and $\mathcal{D}$ respectively given by \eqref{cestpi} and \eqref{diracoper}, has a diagonal action on $\mathbb{H}$ \eqref{hilbert-final}. Hence, it is enough to consider the compactness of $\pi_\nu(a)(1+{D}^2)^{-1/2}:=\pi_\nu(a)T^{-1/2}$ as an operator on $\mathcal{H}_\nu$.\\
Now a simple calculation using \eqref{actionderivation} and \eqref{symfundam} yields
\begin{equation}
(T\varphi)(s)=(1-s^2+\partial_s^2)\varphi(s)\label{cestT}
\end{equation}
for any $\varphi\in\mathcal{H}_\nu$ where the last two terms in \eqref{cestT} correspond to the operator of a one dimensional harmonic oscillator. It follows that the spectrum of $T$ is discrete with eigenvalues $\sim n$, $n\in\mathbb{N}$, with finite multiplicity so that the eigenvalues for $T^{-1}$ vanishes for $n\to\infty$. Notice by the way that $T^{-1}$ alone is therefore compact on $\mathcal{H}_\nu$ and thus $T^{-1/2}$ is also compact. \\
Then, since \eqref{cestT} holds true, the situation is similar to the one covered by \cite[Prop.~5.12]{IMS} which insures that
\begin{equation}
  \pi_\nu(a)(1+{D}^2)^{-1/2}\in\mathcal{L}^{1,\infty}(\mathcal{H}_\nu), 
\end{equation}
for any $a\in\mathbb{A}$ where $\mathcal{L}^{1,\infty}(\mathcal{H}_\nu)$ is the 1st weak Schatten class. Hence $\pi_\nu(a)(1+{D}^2)^{-1/2}$ is compact on $\mathcal{H}_\nu$, implying that $\Pi(a)(1+\mathcal{D}^2)^{-1/2}$ is a compact operator on $\mathbb{H}$ for any $a\in\mathbb{A}$.\\
Finally, upon using the $p,x$ variables to express the unity  \eqref{unities} together with the representations $\pi_\nu$ defined by $(\pi_\nu(a)\varphi)(s)=\int du\ a(u-s,\nu e^{-s})\varphi(u)$, a simple computation yields $\Pi(\bbone)(1+{D}^2)^{-1/2}=(1+{D}^2)^{-1/2}$ which is compact in view of the compactness of $T^{-1/2}$. Therefore, $\Pi(a)(1+\mathcal{D}^2)^{-1/2}$ is compact on $\mathbb{H}$ for any $a\in\tilde{\mathbb{A}}$ which satisfies condition \eqref{cond1bis}.

\section{Causal behaviour of $\kappa$-Minkowski space-time}\label{section3}

We turn now to the exploration of the notion of causality on $\kappa$-Minkowski space-time. The goal here is not to have a complete characterization of the causal sructure, which is still out of sight, but to present a first exploration showing that some specific causality constraints can be extracted using the tool of Lorentzian noncommutative geometry.

\subsection{Introducing causality on a noncommutative quantum space}\label{section31}
In the usual (commutative) framework, the causality is an essential property of any physically relevant theory/model arising as a basic axiom in quantum field theory. The building ingredients are geometric: points, i.e.~events, in a given space-time $M${\footnote{The associated manifold is assumed to be locally compact. Recall that a compact manifold cannot accommodate causality.}}, can eventually be linked together through curves of various types according to the nature of their tangent vectors, e.g. timelike, causal, etc. Then, two points/events are causally related if there exists one causal curve linking them.\\
This geometric description can be translated into an algebraic framework, at least for globally hyperbolic manifolds $M$, the case of interest here. Indeed, trade points and causal curves respectively for pure states and causal functions. Recall that a state of an algebra, $\varphi:\mathbb{A}\to\mathbb{C}$, is a positive linear functional with norm 1. Let $ \mathfrak{S}(\mathbb{A})$ denotes the space of states of $\mathbb{A}=C^\infty_0(M)$. Causal functions are real-valued functions on $M$ which are non-decreasing along every future-directed causal curve, and it is sufficient to restrict to the set of smooth bounded causal functions belonging to ${\tilde{\mathbb{A}}}=C^\infty_b(M)$. The set of these functions has the structure of a convex cone $\mathcal{C}$, $\mathcal{C}\subset\tilde{\mathbb{A}}$, called the causal cone. The technical role of the unitalization ${\tilde{\mathbb{A}}}=C^\infty_b(M)$ is clear here, since if we use the algebra $\mathbb{A}=C^\infty_0(M)$, the cone of causal functions is restricted to constant functions, which are not sufficient to separate points. For globally hyperbolic manifolds $M$, this causal cone determines completely \cite{besnard2} the usual causal structure of $M$ given by
\begin{equation}
     p\preceq q \iff f(p)\le f(q), \ \ \forall f\in \mathcal{C}
\end{equation}
for any $p,q\in M$. \\
A convenient algebraic characterization of the causal cone using the data involved in the Lorentzian spectral triple coding $M$, thus corresponding to a commutative algebra $\mathbb{A}=C^\infty_0(M)$ in \eqref{data1}, has been achieved in \cite{nico1}. Indeed, if we define the causal cone $\mathcal{C}$ as the convex cone of all real-valued functions $f\in\tilde{\mathbb{A}}$ verifying
\begin{equation}
\forall \phi\in\mathcal{H},\\ \langle\phi,\mathcal{J}[D,\pi(f)]\phi\rangle\le 0,
\end{equation}
then \cite[Theorem 7]{nico1} guaranties that, for globally hyperbolic manifolds, the causal structure defined by
\begin{equation}
\forall \omega,\eta\in\mathfrak{S}(\tilde{\mathbb{A}}),\ \omega\preceq \eta\iff \forall f\in\mathcal{C},\ \omega(f)\le\eta(f),
\end{equation}
when restricted to the pure states in $\mathfrak{S}(\mathbb{A})$ is {\it{exactly}} the usual causal structure of $M$.\\

This algebraic framework naturally extends to the noncommutative framework, i.e.~for noncommutative algebra $\mathbb{A}$ in \eqref{data1} to give rise to the following definition of causality which actually defines a reasonable notion of quantum causality:
\begin{itemize}
\item The causal cone $\mathcal{C}$ of a noncommutative Lorentzian spectral triple $\{\mathbb{A},\widetilde{\mathbb{A}},\pi,\mathcal{H} D,\mathcal{J}\}$ is the convex cone of all hermitian elements $f\in\tilde{\mathbb{A}}$ satisfying the condition:\\
\begin{equation}\label{cestcausal}
\forall \phi\in\mathcal{H},\ \ \langle\phi,\mathcal{J}[D,\pi(f)]\phi\rangle\le 0.
\end{equation}
\item The causal relation between (pure) states is defined by the following relation:
\begin{equation}
\forall \omega,\eta\in\mathfrak{S}(\tilde{\mathbb{A}}),\ \omega\preceq \eta\iff \forall f\in\mathcal{C},\ \omega(f)\le\eta(f),\label{cestcausal2}
\end{equation}
which is a well defined partial order relation on all states and if the condition \mbox{$\text{span}_\mathbb{C}(\mathcal{C})=\tilde{\mathbb{A}}$} is respected, ensuring that all states can be separated.
\end{itemize}
Note that the causality should be viewed as a partial order relation between all the states of the unitized algebra, which may be eventually restricted to pure states of the algebra. This definition of quantum causality has been used in past works \cite{nico2,nicojc,nico22,nico-eck} and is the notion of causality which will be examined in the subsequent analysis. \\

Coming back to the case of $\kappa$-Minkowski space-time described above, one easily infers that the relevant causal cone involves the functions $f$ of the group algebra $\mathbb{A}_x$ (or $\mathbb{A}_{px}$ according to the chosen pair of variables) such that 
\begin{equation}
    \bigoplus_{\nu=+,0,-} \left(\begin{matrix} \pi_\nu(\partial_+ f) & 0\\0 &  \pi_\nu(\partial_- f)\end{matrix}\right) \geq 0,\label{conealgeb}
\end{equation}
i.e. \eqref{conealgeb} is a positive operator, where \eqref{diracoper}, \eqref{symfundam} and\eqref{cestcausal} have been used.\\

In order to apply the notion of causality, we need to define the noncommutative equivalent of the notion of point, i.e.~pure states in $\mathfrak{S}({\mathbb{A}})$. The complete set of pure states of $\kappa$-Minkowski space-time is unknown, but an interesting set of pure states can be readily determined  by considering the following family of vector states
\begin{equation}\label{purestates}
\varphi_\pm^\Phi : \mathbb{A} \rightarrow \setC,\ \ \varphi_\pm^\Phi(f)= \scal{\Phi,\pi_\pm(f) \Phi}
\end{equation}
for any $\Phi\in\mathcal{H}_\pm$ with $||\Phi||=1$. Any state belonging to the family \eqref{purestates} is a pure state. To see that, one observes that irreducibility of $(\pi_{U\pm},\mathcal{H}_\pm)$ \eqref{gelf1} is equivalent to the irreducibility of $(\pi_\pm, \mathcal{H}_\pm)$ \eqref{decadix}. Then, any 
non-zero $\Phi\in\mathcal{H}_\pm$ is cyclic for $\mathcal{H}_\pm$. But any vector state of the form $\langle v,\pi(f)v\rangle$ where $f\in\mathcal{A}$, $\mathcal{A}$ some algebra, and $v\in H$, $H$ some Hilbert space, is a cyclic vector for an irreducible representation $(\pi,H)$ of the algebra is a pure state of $\mathcal{A}$. The statement follows.\\

Using \eqref{reprerpnu} and \eqref{purestates}, one easily obtains the following explicit formulation of pure states applied on $\mathbb{A}_{x}$
\begin{equation}
  \varphi_\pm^\Phi(f)  = \tfrac{1}{2\pi} \iiint \dd s \dd u \dd v\; f(v, \pm e^{-s}) \,e^{-i v (u-s)}\, \overline\Phi(s)\Phi(u),\label{stateone}
\end{equation}
or alternatively on $\mathbb{A}_{px}$
\begin{equation}
  \varphi_\pm^\Phi(\tilde{f})=\int\int\dd u \dd s\ 
  \tilde{f}(u-s,\pm e^{-s})\overline\Phi(s)\Phi(u),\label{statetwo}
\end{equation}
for any $\Phi\in\mathcal{H}_\pm$.\\

At this stage, two comments are in order.
\begin{itemize}
    \item We only have here a subset of the total set of pure states, however since those states correspond to irreducible unitary representations, the specific interest of this subset is given by Gel'fand--Raikov theorem.
    \item The above states \eqref{stateone}, \eqref{statetwo} must be extended so as to be well defined on $\tilde{\mathbb{A}}$. This can be consistently achieved by further restricting the vectors $\Phi\in\mathcal{H}_\pm$ to smooth compactly supported functions in $\mathcal{H}_\pm$. This restriction will be assumed in the rest of the discussion.
    \item We can see that the considered pure states are divided into two separated subsets defined by $+$ and $-$ and the only difference separating those subsets is the sign of the second variable of function to which the state applies. Hence the causal structure is completely symmetric, and results concerning the minus representation $\pi_-$ can be easily recovered from the results concerning the plus representation $\pi_+$ from a simple substitution $e^{-s} \rightarrow -e^{-s}$. Hence from now we will restrict most of the time our analyse to the $\pi_+$ side only when the $\pi_-$ side can be trivially recovered.
\end{itemize}
Now from the condition \eqref{cestcausal2}, one realizes that the causality relation between two states is
\begin{equation}
   \varphi_\pm^{\Phi_1} \preceq \varphi_\pm^{\Phi_2} \iff \forall f\in\C_x,\ \ 
  \varphi_\pm^{\Phi_1}(f) \leq \varphi_\pm^{\Phi_2}(f)  ,
\end{equation}
where in obvious notations $\C_x $ is the causal cone defined by \eqref{conealgeb} expressed in the $x$'s variables. These conditions for the existence of a causal evolution from the state $\Phi_1$ to the state $ \Phi_2$ can be easily rewritten as
\begin{equation}\label{evolstatesalpbeta}
 \iiint \dd s \dd u \dd v\; f(v,  \pm e^{-s}) \,e^{-i v (u-s)}\, \left[ \overline\Phi_2(s)\Phi_2(u) - \overline\Phi_1(s)\Phi_1(u) \right] \geq 0,\quad \forall f\in\C_x
\end{equation}
or alternatively in the mixed variable formalism
\begin{equation}\label{evolstatesabeta}
 \iint \dd s \dd u \fabeta(u-s,  \pm e^{-s}) \, \left[ \overline\Phi_2(s)\Phi_2(u) - \overline\Phi_1(s)\Phi_1(u) \right] \geq 0,\quad \forall \fabeta\in\C_{px}.
\end{equation}
\vspace{0.2cm}

\subsection{A sufficient condition: the phase-momentum transport}\label{section32}
In the following, we will derive a sufficient condition generating causal evolutions of states and we will show that phase-momentum transport is a particular solution, hence showing a particular non-trivial possibility of causal evolution on $\kappa$-Minkowski space-time.

We will use here the formalism expressed in term of the mixed variables $(p_0, x_1)$ corresponding to the algebra $\mathbb{A}_{px}$ and its preferred unitalization $\widetilde{\mathbb{A}}_{px}$ and restrict our computations to the $\pi_+$ side of the set of pure states. Using the second fundamental theorem of calculus, the causal constraint \eqref{evolstatesabeta} becomes 
\begin{equation}\label{evolstatescont}
 \iint \dd s \dd u \; \fabeta(u-s,  e^{-s})\, \dv{}{t}\left(  \overline\Phi_t(s)\Phi_t(u) \right) \geq 0,\quad \forall \falpbeta\in\C_{px}
\end{equation}
where $\Phi_t$ represents a step by step evolution in $\H_+$ from $\Phi_1$ to $\Phi_2$ under a continuous parameter $t\in[1,2]$, hence corresponding to a continuous evolution from the state $\varphi_+^{\Phi_1} $ to the state $\varphi_+^{\Phi_2} $.

We need to identify the causal functions in the set $\C_{px}$ which are all functions in $\widetilde{\mathbb{A}}_{px}$ respecting the semi-definite positive constraint \eqref{conealgeb}. Note that we will see in Section \ref{section33} that this set is non-empty. Since this constraint is fully diagonal, a function $\fabeta\in\widetilde{\mathbb{A}}_{px}$ is in $\C_{px}$ if and  only if
\begin{equation}
 \scal{\psi,\pi_\nu(\partial_\pm f)  \psi} \geq 0,\quad\forall \psi \in L^2(\setR)\text{ and } \nu=+,0,-.
\end{equation}
Using the representations in the mixed variables formalism \eqref{reprerpmixed}, 
and the derivations $\partial_\pm$ which become in this formalism
\begin{equation}\partial_\pm ( \fabeta)(p_0,x_1) = (D_0 \pm D_1) ( \fabeta)(p_0,x_1)  = (i p_0 \pm x_1 \partial_1) ( \fabeta)( p_0,x_1), \end{equation}
we get
\begin{align}
\scal{\psi,\pi_\nu(\partial_\pm f)  \psi} &=  \int \dd s  \; \overline\psi(s) \left[ \pi_\nu(\partial_\pm f)  \psi \right](s)\\
 &=  \int \dd s  \; \overline\psi(s)  \int \dd u\; \partial_\pm\fabeta(u-s, \nu e^{-s})\, \phi(u) \\
 &=  \iint \dd s \dd u \; \left[ i(u-s) \fabeta(u-s, \nu  e^{-s}) \pm \nu e^{-s} \partial_1\fabeta(u-s, \nu e^{-s})  \right] \,  \overline\psi(s)\psi(u).
\end{align}
We must note here that, since we are only considering states $\varphi_+^\Phi $ corresponding to the representation $\pi_+$, the constraint \eqref{evolstatescont} only dependent on the positive part of the second variable of causal functions $\fabeta(u-s,  e^{-s})$, so we can ignore all $\nu=0,-$ and get the following inequalities for an arbitrary causal function  $\fabeta\in\C_{px}$ and $\psi \in L^2(\setR)$
\begin{equation}\label{causalineq}
\iint \dd s \dd u \; \left[ i(u-s) \fabeta(u-s,   e^{-s}) \pm  e^{-s} \partial_1\fabeta(u-s,  e^{-s})  \right] \,  \overline\psi(s)\psi(u) \geq 0.
\end{equation}
Since \eqref{causalineq} is valid for every $\psi \in L^2(\setR)$, it is valid for every $\psi \in \C^\infty_c(\setR)$ (recall that we have restricted our set of states to pure states defined from smooth compactly supported functions $\Phi$).

Let us consider the function 
\begin{equation}
F(s,u) =  \fabeta(u-s, e^{-s}) \overline\psi(s)\psi(u)
\end{equation}
for $\psi \in \C^\infty_c(\setR)$. Then $F$ is smooth with compact support, therefore its limits $s,u\rightarrow\infty$ are null and Stokes' theorem provides 
\begin{equation}
\iint \dd s \dd u \ \vec\nabla \cdot F(s,u) \vec 1 = \iint \dd s \dd u (\pd F s + \pd F u)  = 0.
\end{equation}
However,
\begin{equation}
\pd F s =  -\partial_0\fabeta(u-s, e^{-s}) \overline\psi(s)\psi(u)  - e^{-s} \partial_1\fabeta(u-s, e^{-s}) \overline\psi(s)\psi(u)  +  \fabeta(u-s, e^{-s}) \overline\psi^\prime(s)\psi(u)
\end{equation}
\begin{equation}
\pd F u =   \partial_0\fabeta(u-s, e^{-s}) \overline\psi(s)\psi(u)  + \falpbeta(u-s, e^{-s}) \overline\psi(s)\psi^\prime(u)
\end{equation}
which gives
\begin{equation}\label{calculcas1} \iint \dd s \dd u \,e^{-s} \partial_1\fabeta(u-s, e^{-s}) \overline\psi(s)\psi(u) =  \iint \dd s \dd u   \fabeta(u-s, e^{-s}) (\overline\psi^\prime(s)\psi(u) +  \overline\psi(s)\psi^\prime(u))   \end{equation}
and the inequalities \eqref{causalineq} holding for every $\fabeta\in\C_{px}$ and $\psi \in \C^\infty_c(\setR)$ become
\begin{equation}
 \iint \dd s \dd u \;   \fabeta(u-s, e^{-s}) \left[ i(u-s)  \overline\psi(s)\psi(u) \pm\prt{ \overline\psi^\prime(s)\psi(u) + \overline\psi(s)\psi^\prime(u) } \right] \geq 0
\end{equation}
which can be turned using a convex combination of the $+$ and $-$ formulations into a continuous set of inequalities  
\begin{equation}\label{condcaustransf}
 \iint \dd s \dd u \;   \fabeta(u-s, e^{-s}) \left[ i(u-s)  \overline\psi(s)\psi(u) + \alpha\prt{ \overline\psi^\prime(s)\psi(u) + \overline\psi(s)\psi^\prime(u) } \right] \geq 0
\end{equation}
with $\alpha \in [-1,1]$.

By comparing \eqref{evolstatescont} and \eqref{condcaustransf}, we get a sufficient condition for a causal evolution between two states $\varphi_+^{\Phi_1} \preceq \varphi_+^{\Phi_2} $ represented by the continuous evolution of $\Phi_t$ if there exists $\forall t\in [1,2]$ a function $\psi_t \in \C^\infty_c(\setR)$ and $\alpha_t \in [-1,1]$ such that
\begin{equation}\label{CSgen}
 \dv{}{t}\left(  \overline\Phi_t(s)\Phi_t(u) \right) =    i(u-s)  \overline\psi_t(s)\psi_t(u) + \alpha_t \prt{ \overline\psi_t^\prime(s)\psi_t(u) + \overline\psi_t(s)\psi_t^\prime(u) }.
\end{equation}
\vspace{0.2cm}

The sufficient condition \eqref{CSgen} accepts of course trivial solutions given by $\Phi_t$ constant on $t$ and $\psi_t=0$, which correspond to stationary states. While it is difficult to catalog all possibilities, we can highlights a simple and interesting no trivial set of solutions. Indeed, if we equalize $\psi_t = \Phi_t$, the condition \eqref{CSgen} becomes
\begin{align} \dv{}{t}\left(  \overline\Phi_t(s)\Phi_t(u) \right)
& =  \overline{\left( \dv{}{t} \Phi_t(s) \right )} \Phi_t(u)  + \overline\Phi_t(s) \left( \dv{}{t}  \Phi_t(u)  \right ) \\
&=    \overline{is \Phi_t(s)} \Phi_t(u) + \overline{ \Phi_t(s)} iu \Phi_t(u) + \alpha_t \prt{ \overline\Phi_t^\prime(s)\Phi_t(u) + \overline\Phi_t(s)\Phi_t^\prime(u) } 
\end{align}
which is solved by 
\begin{equation}
\dv{}{t}  \Phi_t(u) -  \alpha_t \Phi_t^\prime(u)   =  iu \Phi_t(u).
\end{equation}
For $\alpha_t=\alpha$ constant, this equation is a transport equation whose general solution can be derived using the method of characteristics and is given by
\begin{equation}
 \Phi_t(u) = \Phi_0(u+\alpha t)e^{itu} e^{\alpha i \frac{t^2}2}. 
 \end{equation}
 Since only the product $ \overline{\Phi_t(s)} \Phi_t(u)$ appears in the definition of a state $\varphi_+^{\Phi_t} $, the coefficient $e^{\alpha i \frac{t^2}2}$ can be ignored, and we can say that a state $\varphi_+^{\Phi_t} $ can causally evolve from a state  $\varphi_+^{\Phi_1} $ to a state  $\varphi_+^{\Phi_2} $ if $\varphi_+^{\Phi_t} $ is defined with the following vector
\begin{equation} \label{phase-momentum}
\Phi_t(u) = \Phi_0(u+\alpha t)e^{itu}  
 \end{equation}
where $\Phi_0 \in \C^\infty_c(\setR)$ and $\alpha \in [-1,1]$. This evolution represents a $\alpha t$ translation at the level of $\Phi_t$ simultaneously to a $t$ translation at the level of $\mathcal{F}\Phi_t$, hence can be interpreted as a "phase-momentum transport" within the $\kappa$-Minkowski space-time.\\
\vspace{0.2cm}

\subsection{A necessary condition: the quantum causality constraint}\label{section33}
In the previous section, we have shown that the causal constraint \eqref{evolstatesalpbeta}-\eqref{evolstatesabeta} can be turned into a sufficient condition which accepts some specific solutions, hence causal evolution is allowed inside the $\kappa$-Minkowski space-time. In this section, we will take the opposite path, showing that there exists a necessary condition and that such a condition prevents some causal relations inside the $\kappa$-Minkowski space-time. Thereby, while having a complete characterisation of allowed and forbidden causal relations on this quantum space-time is for the moment out of scope, we prove that the causal constraint implies at the same time valid possibilities and restrictions concerning  causality.

The formalism in space-time variables $(x_0,x_1)$ is easier to handle concerning the study of the necessary condition. In order to extract a counterexample $\varphi_+^{\Phi_1} \npreceq \varphi_+^{\Phi_2} $ from the causal constraint \eqref{evolstatesalpbeta}, it is sufficient to find a specific causal function $f \in \C_x$ not respecting the inequality
\begin{equation}\label{strictin}
 \iiint \dd s \dd u \dd v\; f(v,  e^{-s}) \,e^{-i v (u-s)}\, \left[ \overline\Phi_2(s)\Phi_2(u) - \overline\Phi_1(s)\Phi_1(u) \right] \geq 0.
\end{equation}

An easy way to construct some causal functions is the use of a "split" formalism $f(x_0,x_1)=h(x_0)+g(x_1)$. Indeed, the application of a state on such a split function gives
\begin{align}
 \varphi_\pm^\Phi(h+g)  &= \tfrac{1}{2\pi} \iiint\dd s \dd u \dd v\; [h(v) + g (\pm e^{-s})] \,e^{-i v (u-s)}\, \overline\Phi(s)\Phi(u)\\
 &= \tfrac{1}{2\pi} \iiint \dd s \dd u \dd v\; h(v) \,e^{-i v (u-s)}\, \overline\Phi(s)\Phi(u)
\\  &\qquad+ \tfrac{1}{2\pi} \iiint \dd s \dd u \dd v\; g (\pm e^{-s}) \,e^{-i v (u-s)}\, \overline\Phi(s)\Phi(u)\\
&=\tfrac{1}{2\pi} \int \dd v\; h(v) \overline{\int \dd s\;e^{-i v s} \Phi(s)}  {\int \dd u\;e^{-i v u} \Phi(u)}\\
&\qquad + \int \dd s\; g (\pm e^{-s}) \overline\Phi(s) \left[ \tfrac{1}{2\pi} \int \dd v\; e^{i v s} \left(    {\int \dd u\;e^{-i v u} \Phi(u)} \right)   \right] \\
& =\tfrac{1}{2\pi} \int \dd v\; h(v) \abs{\mathcal{F}\Phi(v)}^2 + \int\dd s \,g (\pm e^{-s}) \abs{\Phi(s)}^2.
\end{align}
Using the same splitting process, the condition \eqref{conealgeb} becomes $f(x_0,x_1)=h(x_0)+g(x_1) \in\C_x$ if and only if :
\begin{equation}\label{conditioninithg}
 \tfrac{1}{2\pi} \int \dd v\; h^\prime (v) \abs{\mathcal{F}\psi(v)}^2 \pm \int \dd s \, \nu e^{-s} g^\prime (\nu e^{-s}) \abs{\psi(s)}^2 \geq 0,\quad\forall \psi \in L^2(\setR)
\end{equation}
where $\nu=+1,0,-1$.

Due to Plancherel's theorem, the inequality \eqref{conditioninithg} is automatically satisfied by functions  $f(x_0,x_1)=h(x_0)+g(x_1) $ respecting $h^\prime (v) = 1$ and $\abs{e^{-s} g^\prime (\nu e^{-s}) }\leq 1$, hence with $h (x_0) = x_0$ and $g(x_1)$ at most logarithmically increasing. In particular, the first coordinate function $f(x_0,x_1)=x_0$ with $g(x_1)=0$, corresponding to a global time as in \eqref{timefunction}, is a causal function.

In order to reach an interesting necessary condition, we need to construct causal functions leading to the equality case of  \eqref{conditioninithg} either for the plus or the minus sign. Once more we will focus on the $\pi_+$ side (the $\pi_-$ side is similar) and we will only be interested by the behaviour of the causal functions for the positive side of the second variable $x_0\in\setR^+$, hence with $\nu=+1$, with an implicit smooth extension on $\setR$. An easy way would be to take a function $f(x_0,x_1)=h(x_0)+g(x_1) $  constructed such that $h (x_0) = x_0$ and $e^{-s} g^\prime (e^{-s}) = \mp 1$, which is the case for $g(x_1)=\pm\ln(x_1)$. However, such a function cannot be smoothly extended to $\setR$. Therefore we must consider two sequences of functions $g^\pm_\epsilon(x_1)=\pm\ln(x_1+\epsilon)$ defined on $\setR^+$ and easily extendable on $\setR$ such that the inequality \eqref{conditioninithg} is still respected (the exact behaviour of this extension has no implication on \eqref{strictin}). Since we have  $\abs{e^{-s} {g_\epsilon^\pm}^\prime ( e^{-s}) } = \frac{e^{-s} }{e^{-s} +\epsilon}\leq 1$, the functions $f^\pm_\epsilon = h + g^\pm_\epsilon$ are causal, which means in the meantime that the set $\C_x$ is non-empty and contains non-trivial functions.

We can now check those specific causal functions on the phase-momentum transport specific solution \eqref{phase-momentum} but with an arbitrary $\alpha\in\setR$
\begin{equation}
\Phi_t(u) = \Phi_0(u+\alpha t)e^{itu} .
 \end{equation}
 Using the split functions formalism and similar Fourier transforms on \eqref{evolstatesalpbeta}, we get that, in order to have a valid causal evolution $\varphi_+^{\Phi_0} \preceq \varphi_+^{\Phi_t}$, the following inequality should necessarily be respected for every causal function $f=h+g\in\C_x$
\begin{equation}
\tfrac{1}{2\pi} \int \dd v\; h(v) \prt{ \abs{\mathcal{F}\Phi_t(v)}^2 - \abs{\mathcal{F}\Phi_0(v)}^2 } + \int \dd s \,g (e^{-s}) \prt{\abs{\Phi_t(s)}^2 - \abs{\Phi_0(s)}^2} \geq 0.
\end{equation}
We can note that 
 \begin{equation}
\mathcal{F}\Phi_t(u) = \Phi_0(u-t), \quad \abs{\Phi_t(u)} = \abs{\Phi_0(u+\alpha t)}
 \end{equation}
 which leads to
 \begin{equation}
 \tfrac{1}{2\pi}  \int \dd v\; h(v) \prt{ \abs{\mathcal{F}\Phi_0(v-t)}^2 - \abs{\mathcal{F}\Phi_0(v)}^2 } + \int \dd s \,g (e^{-s}) \prt{\abs{\Phi_0(s+\alpha t)}^2 - \abs{\Phi_0(s)}^2} \geq 0
  \end{equation}
and using standard substitutions
 \begin{equation} 
 \tfrac{1}{2\pi}  \int\dd v\; \prt{h(v+t) -h(v) } \abs{\mathcal{F}\Phi_0(v)}^2  + \int \dd s \,\prt{g (e^{-s+\alpha t}) - g (e^{-s}) } \abs{\Phi_0(s)}^2\geq 0.
     \end{equation}
Now using the specific causal functions $f^\pm_\epsilon(x_0,x_1) = h(x_0) + g^\pm_\epsilon(x_1) = x_0 \pm\ln(x_1+\epsilon)$, we obtain the necessary condition
 \begin{equation} 
 \tfrac{1}{2\pi}  \int\dd v\; t\; \abs{\mathcal{F}\Phi_0(v)}^2  \pm \int \dd s \,\prt{\ln(e^{-s+\alpha t}+\epsilon) - \ln (e^{-s}+\epsilon) } \abs{\Phi_0(s)}^2\geq 0
     \end{equation}
which provides when $\epsilon\rightarrow 0$
 \begin{equation} 
 \tfrac{1}{2\pi}  \int\dd v\; t\; \abs{\mathcal{F}\Phi_0(v)}^2  \pm \int \dd s \, \alpha t \;\abs{\Phi_0(s)}^2\geq 0
     \end{equation}
     leading to 
    \begin{equation} 
 t \pm \alpha t \geq 0
     \end{equation}
using Plancherel's theorem.  Hence the causal evolution of the phase-momentum transport specific solution is clearly forbidden for $\abs{\alpha}>1$.

If we apply the same process on two generic states $\varphi_+^{\Phi} \preceq \varphi_+^{\xi} $  with a causal relation, we have  for every causal function in the split form $f = h + g \in \C_x$
\begin{equation}
  \tfrac{1}{2\pi} \int \dd v\; h(v) \prt{ \abs{\mathcal{F}\xi(v)}^2 - \abs{\mathcal{F}\Phi(v)}^2 } + \int \dd s \,g (e^{-s}) \prt{\abs{\xi(s)}^2 - \abs{\Phi(s)}^2} \geq 0
\end{equation}
and if we consider the same specific sequences of functions  $f^\pm_\epsilon(x_0,x_1) = h(x_0) + g^\pm_\epsilon(x_1) = x_0 \pm\ln(x_1+\epsilon)$ and the induced limit $\epsilon\rightarrow 0$ on the inequality, we obtain the following general necessary condition
\begin{equation}\label{quantumform1}
 \tfrac{1}{2\pi}  \int \dd v\; v \abs{\mathcal{F}\xi(v)}^2  -  \tfrac{1}{2\pi} \int \dd v\; v \abs{\mathcal{F}\Phi(v)}^2  \geq  \abs{\int\dd s \; s  \abs{\xi(s)}^2 -  \int\dd s  \;s \abs{\Phi(s)}^2} .
 \end{equation}
 
 We can show that \eqref{quantumform1} possesses an interpretation in term of quantum operators on the space $L^2(\mathbb{R})$. Indeed, if $X$ represent the position operator, then 
 \begin{equation}
 \int\dd s \; s  \abs{\xi(s)}^2 = \expval{X}{\xi}.
 \end{equation}
 Using Fourier transform on derivative and Parseval's theorem, we also get 
 \begin{align}
 \tfrac{1}{2\pi}  \int \dd v\; v \abs{\mathcal{F}\xi(v)}^2 &=    \tfrac{1}{2\pi} \int \dd v\; \overline{\mathcal{F}\xi(v)} v \mathcal{F}\xi(v) \\
 &=   \tfrac{1}{2\pi}  \int \dd v\; \overline{\mathcal{F}\xi(v)} \mathcal{F}{\prt{-i\dv{}{s}\xi}}(v)\\
 & =  \int \dd s\; \overline{\xi(s)} \prt{-i\dv{}{s}}\xi \\
 & = \expval{P}{\xi}
 \end{align}
 where $P$ is the momentum operator. Therefore, the necessary condition  \eqref{quantumform1}  can be interpreted as 
 \begin{equation}
 \expval{P}{\xi} - \expval{P}{\Phi} \geq \abs{\expval{X}{\xi} - \expval{X}{\Phi}}  \end{equation}
  \begin{equation}\lequi \delta \expval P \geq \abs{\delta \expval X} \end{equation}
  where $\delta \expval X$ represent the variation of the expectation value.
  
  This necessary condition is clearly a quantum analogous to the classical speed of light limit. Indeed, on a two dimensional Minkowski spacetime, two events are causally related if and only if variations of the spatial coordinate $\delta x$ and time coordinate $\delta t$ respect the inequality $\delta t \geq \abs{\delta x}$. A similar restriction occurs within the $\kappa$-Minkowski space-time, with the correspondence $X \sim x$, $P \sim t$.


\appendix
\section{The 2-dimensional $\kappa$-Poincar\'e algebra.}\label{apendixA}

It is convenient to make use of the bicrossproduct basis \cite{majid-ruegg}. Thorough this paper, $\Delta:\mathcal{P}_\kappa\otimes\mathcal{P}_\kappa\to\mathcal{P}_\kappa$, $\epsilon:\mathcal{P}_\kappa\to\mathbb{C}$ and ${\bf{S}}:\mathcal{P}_\kappa\to\mathcal{P}_\kappa$ denote respectively the coproduct, counit and antipode, giving to $\mathcal{P}_\kappa$ a Hopf algebra structure. We denote by $P_0,P, N, \mathcal{E},\mathcal{E}^{-1}$, the energy, spatial momentum, boosts and $\mathcal{E}:=e^{-P_0/\kappa}$. These are the generators of the following Lie algebra
\begin{equation}
 [P_0,\mathcal{E}]=[P,P_0]=[P,\mathcal{E}]=0,\ [N,\mathcal{E}]=\frac{i}{\kappa}P\mathcal{E},\ [N,P]=-\frac{i\kappa}{2}(1-\mathcal{E}^{2})+\frac{i}{2\kappa}P^2\label{poinc2}.
\end{equation}
The relations defining the Hopf algebra structure of $\mathcal{P}_\kappa$
\begin{align}
\Delta P_0&=P_0\otimes\bbone+\bbone\otimes P_0,\ \Delta P=P\otimes\bbone+\mathcal{E}\otimes P,\ \Delta \mathcal{E}=\mathcal{E}\otimes\mathcal{E}\label{hopf1},\\
\Delta N&=N\otimes \bbone+\mathcal{E}\otimes N,\label{hopf2}\\
\epsilon(P_0)&=\epsilon(P)=\epsilon(N)=0,\  \epsilon(\mathcal{E})=1\label{hopf3},\\
{\bf{S}}(P_0)&=-P_0,\ {\bf{S}}(\mathcal{E})=\mathcal{E}^{-1},\  {\bf{S}}(P)=-\mathcal{E}^{-1}P,\ {\bf{S}}(N)=-\mathcal{E}^{-1}N.
\end{align}
We denote by $\mathcal{T}_\kappa$ the deformed translation algebra which is the dual of $\mathcal{M}_\kappa$. It is a Hopf subalgebra of $\mathcal{P}_\kappa$ generated by $P_\mu$, $\mathcal{E}$, with involutive structure thanks to $P_\mu^\dag=P_\mu$, $\mathcal{E}^\dag=\mathcal{E}$. One has for any $t$ in $\mathcal{T}_\kappa$ and any $f\in\mathcal{M}_\kappa$ the useful relation:
\begin{equation}
(t\triangleright f)^\dag={\bf{S}}(t)^\dag\triangleright f^\dag,\label{pairing-involution}.
\end{equation}
From this follow
\begin{equation}
(P_0\triangleright f)^\dag=-P_0\triangleright(f^\dag),\ (P\triangleright f)^\dag=-\mathcal{E}^{-1}P\triangleright(f^\dag),\ (\mathcal{E}\triangleright f)^\dag=\mathcal{E}^{-1}\triangleright(f^\dag)\label{dag-hopfoperat}.
\end{equation}
The action of $\mathcal{T}_\kappa$ on $\mathcal{M}_\kappa$ is 
\begin{equation}
(\mathcal{E}\triangleright f)(x)=f(x_0+\frac{i}{\kappa},\vec{x}),\ \ 
(P_\mu\triangleright f)(x)=-i(\partial_\mu f)(x)\label{left-module1}.\\
\end{equation}

\section*{References} 
\bibliography{kappa}

\end{document}